\documentclass[]{spie}  

\usepackage{amsmath,amsfonts,amssymb}
\usepackage{subcaption}
\usepackage{graphicx}
\usepackage{cite} 
\usepackage{times}
\usepackage{epsfig}
\usepackage{amsmath}
\usepackage{nccmath}
\usepackage{amssymb}
\usepackage{mwe}
\usepackage{acro}
\usepackage{amssymb}
\usepackage{xcolor,colortbl}
\usepackage{tabularx}
\usepackage{relsize}
\usepackage{pifont}
\usepackage{booktabs} 
\usepackage{multirow}
\usepackage{multicol}
\usepackage{adjustbox}
\usepackage{float}
\usepackage{graphicx}
\usepackage{makecell}
\usepackage{tabu}
\usepackage[colorlinks=true, allcolors=blue]{hyperref}
\usepackage[capitalize]{cleveref}

\usepackage{algorithm}
\usepackage{algorithmic}

\title{Deep Learning-Based Open Source Toolkit for Eosinophil Detection in Pediatric Eosinophilic Esophagitis}

\author[a]{Juming Xiong}
\author[a]{Yilin Liu}
\author[a]{Ruining Deng}
\author[b]{Regina N Tyree}
\author[c]{Hernan Correa}
\author[b]{Girish Hiremath}
\author[c]{Yaohong Wang}
\author[a,c,d]{Yuankai Huo}

\affil[a]{Department of Computer Science, Vanderbilt University, Nashville, TN, USA}
\affil[b]{Division of Pediatric Gastroenterology, Hepatology, and Nutrition, Vanderbilt University Medical Center, Nashville, TN, USA}
\affil[c]{Department of Pathology, Microbiology and Immunology, Vanderbilt University Medical Center, Nashville, TN, USA}
\affil[d]{Department of Electrical and Computer Engineering, Vanderbilt University, Nashville, TN, USA}

\authorinfo{Corresponding author: Yuankai Huo: E-mail: yuankai.huo@vanderbilt.edu}

\begin{document} 
\maketitle

\begin{abstract}
Eosinophilic Esophagitis (EoE) is a chronic, immune/antigen-mediated esophageal disease, characterized by symptoms related to esophageal dysfunction and histological evidence of eosinophil-dominant inflammation. Owing to the intricate microscopic representation of EoE in imaging, current methodologies which depend on manual identification are not only labor-intensive but also prone to inaccuracies. In this study, we develop an open-source toolkit, named Open-EoE, to perform end-to-end whole slide image (WSI) level eosinophil (Eos) detection using one line of command via Docker. Specifically, the toolkit supports three state-of-the-art deep learning-based object detection models. Furthermore, Open-EoE further optimizes the performance by implementing an ensemble learning strategy, and enhancing the precision and reliability of our results. The experimental results demonstrated that the Open-EoE toolkit can efficiently detect Eos on a testing set with 289 WSIs. At the widely accepted threshold of $\geq$ 15 Eos per high power field (HPF) for diagnosing EoE, the Open-EoE achieved an accuracy of 91\%, showing decent consistency with pathologist evaluations. This suggests a promising avenue for integrating machine learning methodologies into the diagnostic process for EoE. The docker and source code has been made publicly available at \url{https://github.com/hrlblab/Open-EoE}.

\end{abstract}


\section{INTRODUCTION}
\label{sec:intro}  
Eosinophilic esophagitis is a chronic disease of the esophagus initiated by an immune responses to food and/or aero-antigens. Clinically, it manifests as symptoms associated with esophageal dysfunction, and its main histological characteristic is an eosinophil-predominant inflammation wherein $\geq$ 15 eosinophils (Eos) (peak count) per high power field (HPF) is considered as diagnostic of EoE~\cite{LIACOURAS20113}. The current approach of manual identification of the number of Eos per HPF within a whole slide image (WSI) by a pathologist is time-consuming, labor intensive, and requires substantial attention to detail. Furthermore, this process is also prone to inconsistencies in the identification techniques and intra- and inter-pathologist variability since the Eos are distributed in small and unevenly distributed patches across the esophageal biopsies. These limitations are critical barriers towards the standardization and reliability of diagnosis of EoE.

\begin{figure*}[t]
\begin{center}
\includegraphics[width=1\linewidth]{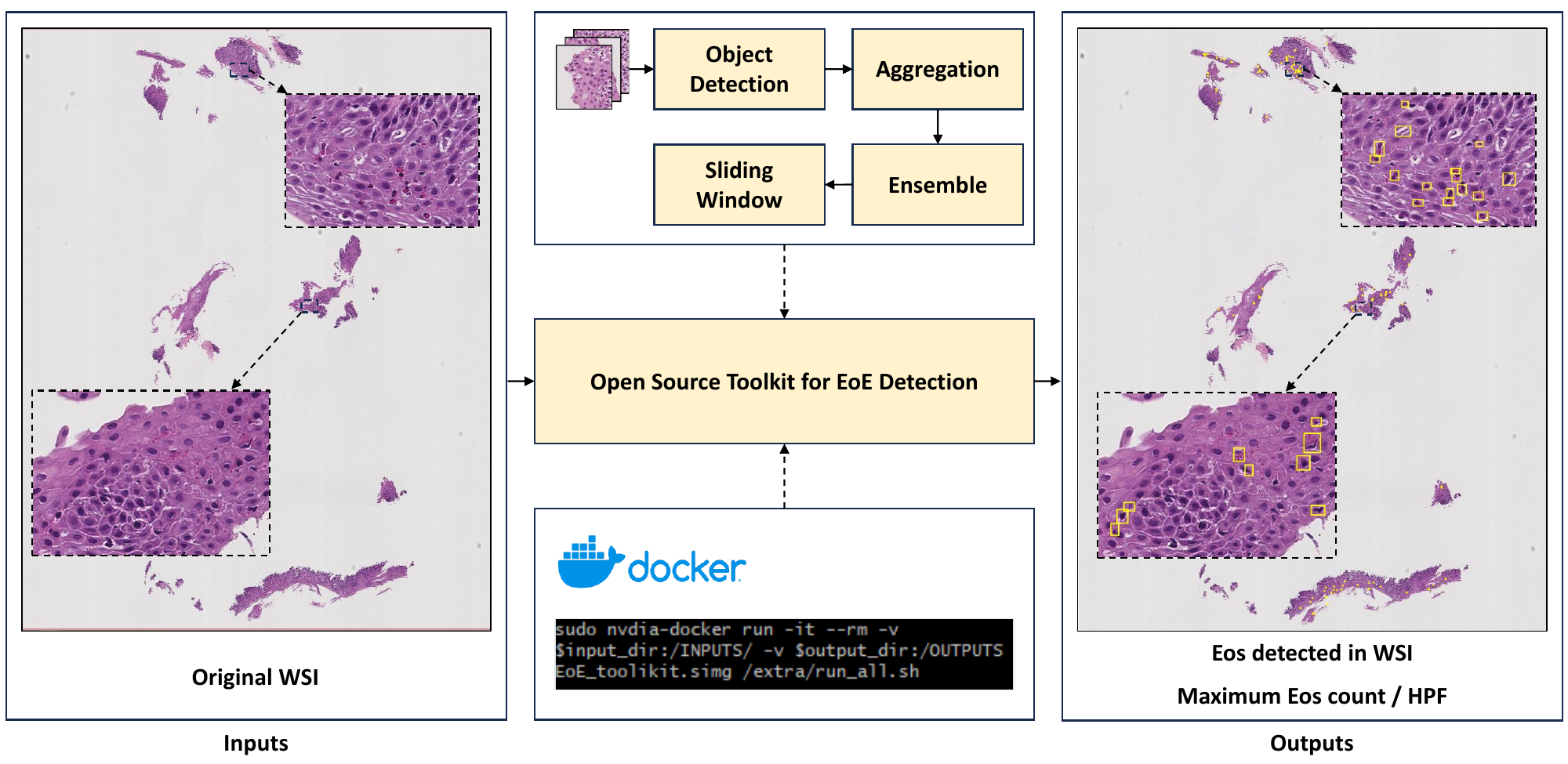}
\end{center}
\caption{This figure shows the overview of the Open-EoE Toolkit. The inputs are original WSIs at 40$\times$ magnification, while the outputs are the maximum Eos count number and the Eos detected bounding boxes that can ovelay on the original WSIs.}
\label{fig:overview}
\end{figure*}

With the popularity of medical digital imaging and the exponential growth of computing power, deep learning, especially convolutional neural network (CNN), has become a key technology for medical image analysis. It also performs well in tasks such as classification and segmentation of histological WSIs~\cite{LITJENS201760,KORBAR201730}. In this study, we propose an efficient deep learning-based toolkit, named Open-EoE for detecting the number of Eos in WSIs as shown in Fig.~\ref{fig:overview}. This computational approach incorporates three prevalent deep learning models for object detection, Faster R-CNN~\cite{Ren_2017}, Mask R-CNN~\cite{He2017MaskR}, and CenterNet~\cite{zhou2019objects}, to identify Eos. It systematically analyzes WSIs, returning the peak count of Eos under each HPF, thereby providing a precise quantitative assessment that can be instrumental in the diagnosis of EoE.

\section{Method}
\begin{figure*}[t]
\begin{center}
\includegraphics[width=1\linewidth]{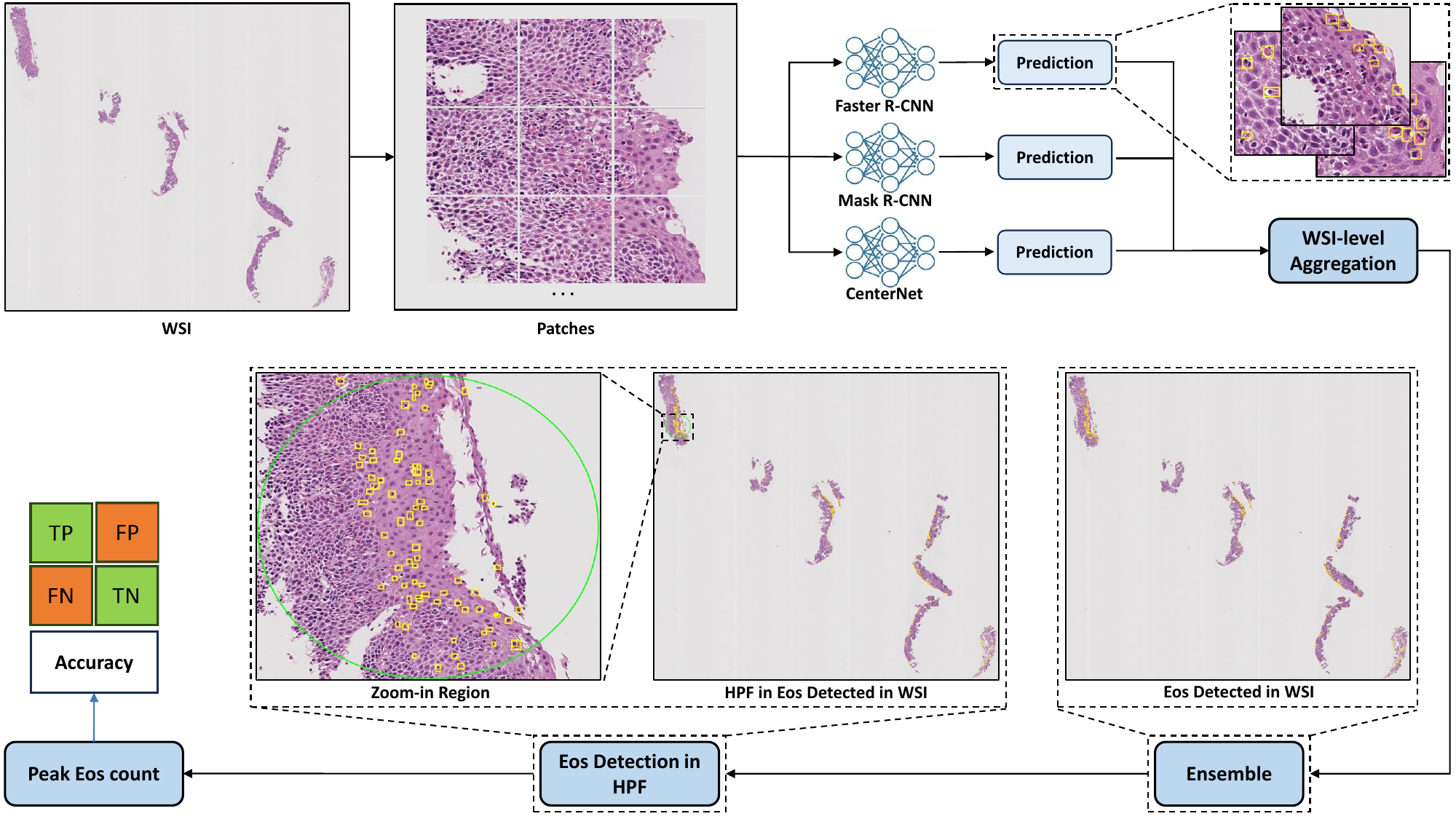}
\end{center}
\caption{This figure shows the detailed pipline of the Open-EoE toolkit. First, the WSI is cut to patches. Second, the patch images are loaded in three deep learning models for training or prediction. Third, the results predicted by each model are aggregated to WSI-level. Fourth, the results of three WSI-level are ensembled. Finally, the maximum value of Eos in HPF is determined and compare with the clinical count.}
\label{fig:Method}
\end{figure*}

This deep learning-based open source toolkit shown in Fig.~\ref{fig:Method} has four major steps: (1) Preprocessing, (2) Object Detection, (3) WSI-level results aggregation, and (4) ensemble learning.

\subsection{Preprocessing}
The annotation of the images was carried out with the collaboration of pathologists from Vanderbilt University Medical Center. We employed Qupath, a renowned open-source software tailored for digital pathology and WSI analysis~\cite{bankhead2017qupath}, for the annotation of these WSIs at 40$\times$ magnification. Once the annotation was finalized, we utilized a built-in script to export all the annotations as JSON files. These files were then converted into the COCO format for model training. COCO is a well-accepted image dataset format extensively employed in the field of computer vision, specifically for functions such as object detection, and segmentation. Every patch measure 512$\times$512 pixels at its original resolution and for underfills, the pad with the white pixel has a value of 255. Each patch aligns precisely with the annotation. Both the patch image and corresponding annotation serve as inputs for training the model. The entire WSI requiring prediction is subdivided into patches using a consistent approach. Each patch is labeled with the coordinates of its upper-left corner vertex, and subsequently employed as an individual input for the model's prediction.

\subsection{Object Detection}
Faster R-CNN~\cite{Ren_2017} stands out as a prominent model for object detection, comprising a Region Proposal Network (RPN) for generating object location proposals, ROI Pooling to resize these regions, and fully connected layers for object class determination and bounding box refinement. Renowned for its high accuracy, Faster R-CNN represents a substantial training enhancement over earlier models like Fast R -CNN~\cite{DBLP:journals/corr/Girshick15}.
Mask R-CNN~\cite{He2017MaskR}, an extension of Faster R-CNN, introduces a parallel branch for predicting object masks, enabling instance segmentation. This model can discern objects and classify each pixel within the object, achieving state-of-the-art performance in both object detection and instance segmentation tasks. CenterNet~\cite{zhou2019objects} presents an innovative approach by detecting objects as points, identifying the center of each object, and predicting properties like size, shape, and class. This streamlined method offers robust detection performance through a simplified architecture. For the implementation of these three networks, we utilized MMDetection~\cite{mmdetection}, an open-source platform tailored for object detection and segmentation.

\subsection{WSI-level results aggregation}

Upon acquiring predictions for bounding boxes within each patch of the entire WSI using these models, we translate these outcomes back to their original coordinates within the overarching WSI. This restorative process entails employing the patch names, which encode the coordinates of each patch's upper-left vertex. Through this information, we meticulously realign the predicted bounding boxes to their accurate positions in the overall slide image. Ultimately, the consolidated outcomes are stored as three JSON files, corresponding to the three deep learning models, for subsequent ensemble learning. Additionally, three GEOJSON files are generated in a format compatible with Qupath, facilitating the observation of detection results at the original resolution using Qupath.

\subsection{Ensemble learning}
We consolidate predictions from the three previously mentioned models at the WSI level through a rigorous voting mechanism. In this methodology, when two or more models concur on an object prediction, it is deemed a valid prediction. Subsequently, we employ non-maximum suppression to the amalgamated bounding boxes, discerning the most accurate prediction outcomes. The conclusive outcomes are then stored in both a JSON file, for counting the peak value of Eos in HPF and a GEOJSON file, for visualization.

\section{Data and Experiments}

\subsection{Data}
Initially, we extracted 10,953 image patches from 50 WSIs sourced from the PediatricEoE dataset, utilizing a 40$\times$ objective lens. Additionally, our dataset encompasses 289 WSIs from 183 patients, totaling more than 500,000 images, each sized 512$\times$512 pixels. These images are slated for prediction employing our model. To enhance the model's discernment of Eos characteristics, we established two distinct classes. Specifically, the ``Eos" class is accompanied by 6,921 annotations, while the ``papillae Eos" class comprises 4,032 annotations. For optimal model training and testing, we methodically divided the dataset into three distinct subsets: a training set, a validation set, and a test set. This division adheres to the conventional distribution ratio of 7:1:2, as comprehensively outlined in the provided Table~\ref{table1}.

\begin{table}[ht]
\caption{Overview of Dataset}
\centering
\begin{tabular}{l@{}c@{\ \ }ccccc}
\toprule
  & \textbf{Train} & \textbf{Val} & \textbf{Test} & \textbf{Total} \\
\midrule
\textbf{Eos}           & 4842 & 690 & 1389 & 6921 \\
\textbf{Papillae Eos}\hspace{1em}  & 2789 & 430 & 813 & 4032 \\
\textbf{Total}         & 7631 & 1120 & 2202 & 10953  \\
\bottomrule
\label{table1}
\end{tabular}
\end{table}

In the second experiment, we employ a dataset consisting of 289 WSIs sourced from 183 patients, amounting to a collection of over 500,000 images, each sized at 512$\times$512 pixels. These images are in need of prediction through our model. Our primary goal is to ascertain the highest Eos/HPF within each patient's WSI. In instances where a patient is associated with multiple WSIs, we extract the maximum Eos value from each set of outcomes. Ultimately, we compare these predicted peak Eos values with the observations made by pathologists at Vanderbilt University Medical Center.


\subsection{Experiment}

\subsubsection{Training Detail}
The network underwent training with fine-tuned parameters derived from the validation set, encompassing a learning rate of 0.01, a momentum value of 0.9, and a weight decay of 0.001. The input image resolution was configured as 512$\times$512$\times$3. For dataset augmentation, we utilized a resizing scale of 1.5625, introduced random flipping and shifting ratios set at 0.5, and integrated photometric distortion techniques. Throughout the training phase, a batch size of 8 was employed, and the chosen optimizer was Stochastic Gradient Descent (SGD).

The implementation of the network was carried out using Python version 3.8 and PyTorch version 2.0.1, utilizing CUDA version 11.7 for GPU acceleration. The experiments were performed on an NVIDIA RTX A5000 GPU with 24GB memory, enabling efficient processing and training of the model.

\subsubsection{Maximum Eos count}
In the original resolution, the pixel width is 0.25 $\mu$m. So, we calculated the Eos density (Eos/mm\textsuperscript{2}), and subsequently converted it to an Eos/HPF assuming a HPF size of 0.24 mm\textsuperscript{2}, which is commonly reported in the literature~\cite{RUSIN201750,dellon2007variability}. In the end, we opted for a circular sliding window with a diameter of 550 mm, which is equivalent to 2200 pixels to identify the peak value of Eos density. This circular sliding window employs global sliding with a stride of 10 pixels.

\section{Results}
The results of the three models and ensemble learning are shown in Table~\ref{table2}. In this experiment, ensemble learning outperforms the other three models in terms of each level of Average Precision (AP) value. Additionally, for the Average Precision for small objects (AP$_{(S)}$), ensemble learning again demonstrates superiority with the highest AP value, surpassing Faster-RCNN, Mask-RCNN, and CenterNet.

\begin{figure*}[t]
\begin{center}
\includegraphics[width=1\linewidth]{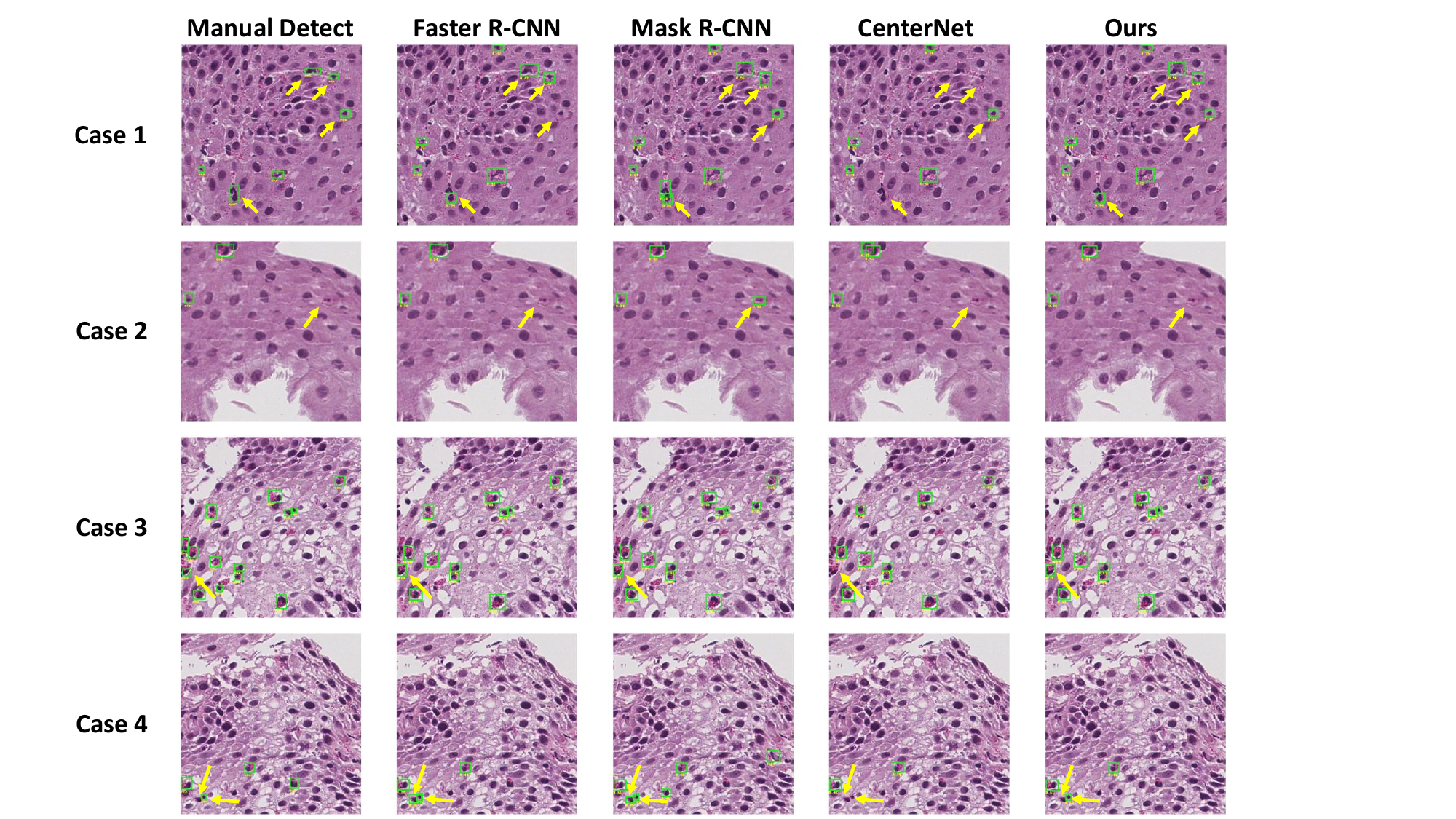}
\end{center}
\caption{This figure displays four cases that the different prediction result from each method compared to the manual detection. Each row indicates a different example patches.Each column indicates Eos detection result from the specific method. The arrows indicate the inconsistent detection results of each method}
\label{fig:result1}
\end{figure*}

In Fig.~\ref{fig:result1}, we compare the results of ensemble learning with those of the other three models against the manually detected bounding boxes. The findings demonstrate that ensemble learning effectively reduces false positives, where non-Eos regions are erroneously predicted as Eos. As a result, the ensemble learning predictions achieve higher accuracy, providing more precise and accurate bounding boxes.

\begin{figure*}[t]
\begin{center}
\includegraphics[width=1\linewidth]{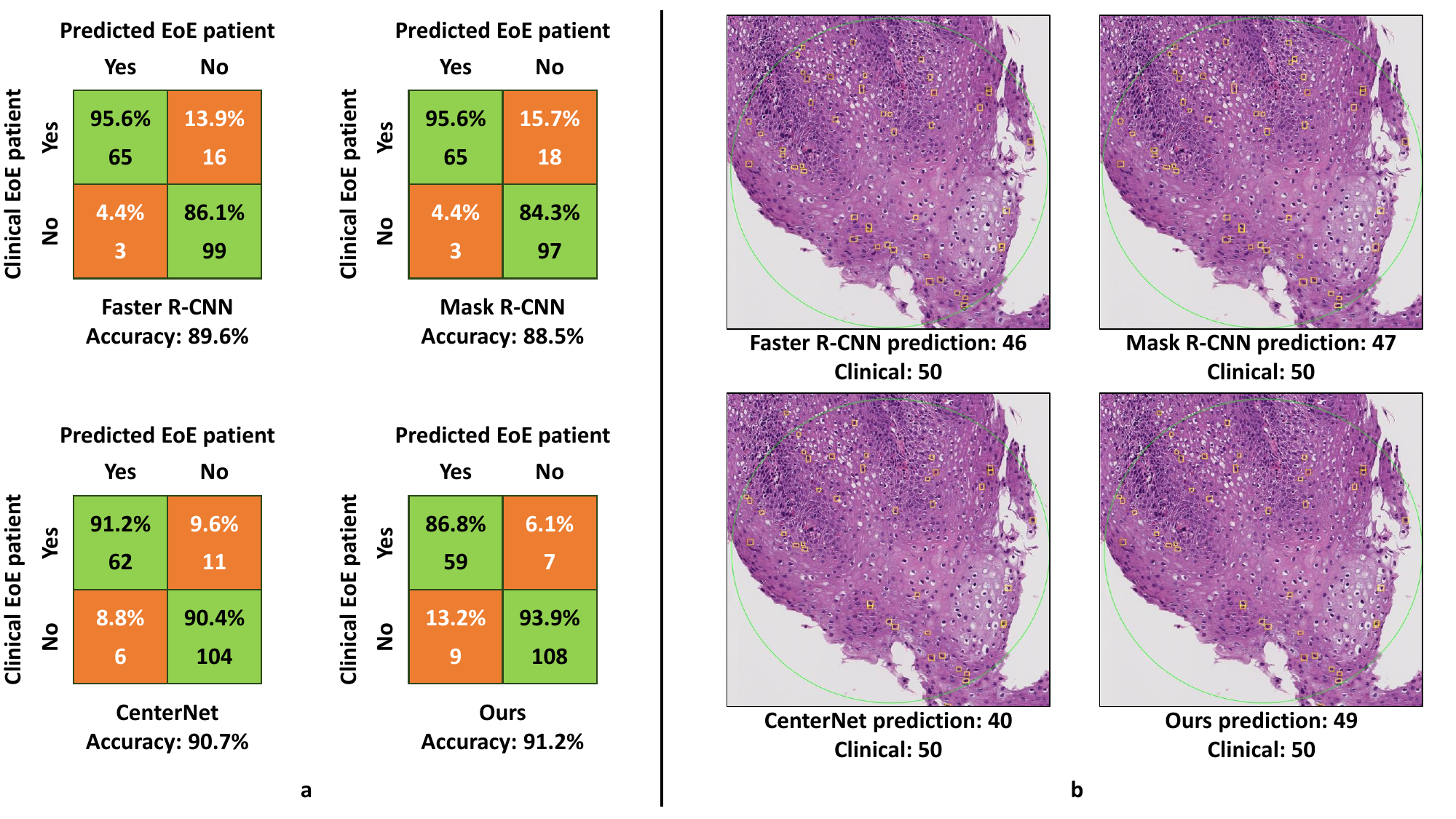}
\end{center}
\caption{Part (a) shows the accuracies of the models compared to the clinical result. True Positive (TP): clinical Eos $\geq 15$ and predicted Eos $\geq 15$; False Negative (FN): clinical Eos $\geq 15$ and predicted Eos $< 15$;False Positive (FP): clinical Eos $< 15$ and predicted Eos $\geq 15$. True Negative (TN): clinical Eos $< 15$ and predicted Eos $< 15$. Part (b) shows a example that the maximum Eos count discovered in HPF using each method, and compared with the Maximum Eos count from clinical records}
\label{fig:result2}
\end{figure*}

We conducted a comparison between the maximum count of Eos detected in WSIs obtained through sliding window techniques and the detection results provided by pathologists at Vanderbilt University Medical Center for 183 patients. As depicted in Fig.~\ref{fig:result2}, all prediction outcomes demonstrate strong performance, with ensemble learning achieving the highest accuracy at 91.2\%. The individual models also display commendable performance with Faster-RCNN scoring 89.6\%, Mask-RCNN achieving 88.5\%, and CenterNet attaining 90.7\%.

\begin{table}[ht]
\caption{Detection performance(\%)}
\centering
\begin{tabular}{l@{}c@{\ \ }cccc}
\toprule
  Models &   $AP$ & $AP_{(50)}$ & $AP_{(75)}$ & $AP_{(S)}$ & $AP_{(M)}$\\
\midrule
Faster R-CNN  & 38.3 & 75.6 & 34.2 & 38.2 & 42.3 \\
Mask R-CNN  & 38.7 & 75.3 & 35.5 & 39.0 & 42.5 \\
CenterNet  & 36.3 & 74.9 & 29.4 & 38.4 & 38.3  \\
Ensemble learning (Ours) \hspace{1em}  & \textbf{39.5} & \textbf{76.3} & \textbf{35.7} & \textbf{39.5} & \textbf{43.7}  \\
\bottomrule
\textbf{\label{table2}}
\end{tabular}
\end{table}

\section{Conclusion}

In this paper, we present an end-to-end deep-learning-based open-source toolkit for Eos detection in Pediatric Eosinophilic Esophagitis. The implementation is made operating system agnostic through the use of Docker, allowing for smooth deployment across different platforms. This toolkit offers a seamless user experience, as it can automatically process a vast number of billion pixel-level WSIs using a single command. Furthermore, it efficiently saves the corresponding results, enhancing user-friendliness. In terms of technical contributions, the key innovation of our research lies in the implementation of an ensemble learning strategy that combines three powerful deep learning models. By leveraging the strengths of multiple models and their collective decision-making, the Open-EoE achieves a higher accuracy for Eos detection compared with using individual models.

\newpage

\section{ACKNOWLEDGMENTS}       
This work has not been submitted for publication or presentation elsewhere. This work is supported in part by NIH R01DK135597(Huo).

\bibliography{main} 

\begin{thebibliography}{10}

\bibitem{LIACOURAS20113}
Liacouras, C.~A., Furuta, G.~T., Hirano, I., Atkins, D., Attwood, S.~E., Bonis,
  P.~A., Burks, A.~W., Chehade, M., Collins, M.~H., Dellon, E.~S., Dohil, R.,
  Falk, G.~W., Gonsalves, N., Gupta, S.~K., Katzka, D.~A., Lucendo, A.~J.,
  Markowitz, J.~E., Noel, R.~J., Odze, R.~D., Putnam, P.~E., Richter, J.~E.,
  Romero, Y., Ruchelli, E., Sampson, H.~A., Schoepfer, A., Shaheen, N.~J.,
  Sicherer, S.~H., Spechler, S., Spergel, J.~M., Straumann, A., Wershil, B.~K.,
  Rothenberg, M.~E., and Aceves, S.~S., ``Eosinophilic esophagitis: Updated
  consensus recommendations for children and adults,'' {\em Journal of Allergy
  and Clinical Immunology}  (2011).

\bibitem{LITJENS201760}
Litjens, G., Kooi, T., Bejnordi, B.~E., Setio, A. A.~A., Ciompi, F.,
  Ghafoorian, M., {van der Laak}, J.~A., {van Ginneken}, B., and Sánchez,
  C.~I., ``A survey on deep learning in medical image analysis,'' {\em Medical
  Image Analysis}  (2017).

\bibitem{KORBAR201730}
Korbar, B., Olofson, A.~M., Miraflor, A.~P., Nicka, C.~M., Suriawinata, M.~A.,
  Torresani, L., Suriawinata, A.~A., and Hassanpour, S., ``Deep learning for
  classification of colorectal polyps on whole-slide images,'' {\em Journal of
  Pathology Informatics}  (2017).

\bibitem{Ren_2017}
Ren, S., He, K., Girshick, R., and Sun, J., ``Faster r-cnn: Towards real-time
  object detection with region proposal networks,'' {\em IEEE Transactions on
  Pattern Analysis and Machine Intelligence}  (2017).

\bibitem{He2017MaskR}
He, K., Gkioxari, G., Doll{\'a}r, P., and Girshick, R.~B., ``Mask r-cnn,'' {\em
  2017 IEEE International Conference on Computer Vision (ICCV)}  (2017).

\bibitem{zhou2019objects}
Zhou, X., Wang, D., and Kr{\"a}henb{\"u}hl, P., ``Objects as points,'' in [{\em
  arXiv preprint arXiv:1904.07850}{\nolinebreak\hspace{0.1em}]},  (2019).

\bibitem{bankhead2017qupath}
Bankhead, P., Loughrey, M.~B., Fern{\'a}ndez, J.~A., Dombrowski, Y., McArt,
  D.~G., Dunne, P.~D., McQuaid, S., Gray, R.~T., Murray, L.~J., Coleman, H.~G.,
  et~al., ``Qupath: Open source software for digital pathology image
  analysis,'' {\em Scientific reports}  (2017).

\bibitem{DBLP:journals/corr/Girshick15}
Girshick, R.~B., ``Fast {R-CNN},'' {\em CoRR}  (2015).

\bibitem{mmdetection}
Chen, K., Wang, J., Pang, J., Cao, Y., Xiong, Y., Li, X., Sun, S., Feng, W.,
  Liu, Z., Xu, J., Zhang, Z., Cheng, D., Zhu, C., Cheng, T., Zhao, Q., Li, B.,
  Lu, X., Zhu, R., Wu, Y., Dai, J., Wang, J., Shi, J., Ouyang, W., Loy, C.~C.,
  and Lin, D., ``{MMDetection}: Open mmlab detection toolbox and benchmark,''
  {\em arXiv preprint arXiv:1906.07155}  (2019).

\bibitem{RUSIN201750}
Rusin, S., Covey, S., Perjar, I., Hollyfield, J., Speck, O., Woodward, K.,
  Woosley, J.~T., and Dellon, E.~S., ``Determination of esophageal eosinophil
  counts and other histologic features of eosinophilic esophagitis by pathology
  trainees is highly accurate,'' {\em Human Pathology}  (2017).

\bibitem{dellon2007variability}
Dellon, E.~S., Aderoju, A., Woosley, J.~T., Sandler, R.~S., and Shaheen, N.~J.,
  ``Variability in diagnostic criteria for eosinophilic esophagitis: a
  systematic review,'' {\em Official journal of the American College of
  Gastroenterology| ACG}  (2007).

\end{thebibliography}
\bibliographystyle{spiebib} 

\end{document}